\documentclass{PoS}
\usepackage{cancel}
\title{Interpretation of the top-quark mass measurements:\\ a theory overview}

\ShortTitle{Interpretation of the 
top-quark mass measurements: a theory overview}

\author{\speaker{Gennaro Corcella}\\
       INFN, Laboratori Nazionali di Frascati\\
        Via E.~Fermi 40, I-00044 Frascati (RM), Italy\\
               E-mail: \email{gennaro.corcella@lnf.infn.it}}

\abstract{I discuss the theoretical interpretation of the top-quark
mass, which is extracted in standard and
alternative measurements at the LHC. In particular, I point out that
the top mass extracted in analyses relying on the use of 
Monte Carlo event generators
must be close to the pole mass 
and review recent work aiming at estimating the theoretical uncertainty.}

\FullConference{8th International Workshop on Top Quark Physics\\
		 14-18 September, 2015\\
		 Ischia, Italy}

\begin{document}
\section{Introduction}
The top quark mass is a fundamental parameter of the
Standard Model, which, together with the $W$ mass,
constrained the Higgs mass even before its discovery.
It also plays a role in the result that
the Standard Model vacuum lies on the border
between stability and metastability regions \cite{degrassi},
which relies on the assumption 
that the world average, 
$m_t=[173.34\pm 0.27{\rm (stat)} \pm 0.71{\rm (syst)}]$~GeV \cite{wave},
is the top pole mass. Any deviation from
this statement may change the conclusions of \cite{degrassi}.

Top-quark mass measurements are performed by comparing experimental
results with theory predictions: the extracted mass must thus be
interpreted as the mass used in the 
calculation.
Standard measurements, based on the template, matrix-element
and ideogram methods (see, e.g., the recent analyses in \cite{atlas1,cms1}),
rely on parton shower generators,
such as HERWIG \cite{herwig6} or PYTHIA \cite{pythia6},
which simulate the hard scattering at leading order (LO),
multiple radiation in the soft or collinear
approximation and are provided with models
for hadronization.
The aMC@NLO \cite{mcnlo} and POWHEG \cite{powheg} codes 
generate the hard process at NLO and are matched to HERWIG or
PYTHIA for showers and hadronization.
In principle, any theoretically well-defined top mass extraction
would need at least a NLO calculation for top-quark production
and decay, including interference effects:
much debate has hence been taking place 
on whether measurements using parton showers and
hadronization models correspond to any top-mass definition.
As will be discussed hereinafter, the measured 
top mass must be close
to the pole mass and work has been lately undertaken
to assess the theoretical uncertainty.
So-called alternative methods use other observables,
such as total cross sections or distribution endpoints, 
which can be compared directly with fixed-order and
possibly resummed QCD calculations, thus allowing a
straightforward interpretation of the extracted mass.

In Section 2, I briefly review the main mass definitions; in
Sections 3 I discuss in more detail 
the interpretation of the top mass 
results; in Section 4 I make some final remarks.

\section{Top mass definitions}
Mass definitions are related to
the subtraction of the ultraviolet divergences in the 
renormalized self energy $\Sigma^R$, 
which, in dimensional regularization, with
$d=4-2\epsilon$, at one loop in QCD, reads:
\begin{equation}\label{sigma}
\Sigma^R\simeq\frac{i\alpha_S}{4\pi}\left\{
\left(\frac{1}{\epsilon}-\gamma+\ln 4\pi+A\right)\cancel p-
\left[4\left(\frac{1}{\epsilon}-\gamma+\ln 4\pi\right)+B\right]m_0\right\}+
i[(Z_2-1)\cancel p-(Z_2Z_m-1)m_0],\end{equation}
where $m_0$ is the bare mass, $Z_2$ and
$Z_m$ the wave-function and mass renormalization constants.
The on-shell renormalization scheme, leading to the pole mass definition,
is defined so that 
$\Sigma^R=0$ and $\partial\Sigma^R/\partial\cancel p=0$ for $\cancel p=0$;
the $\overline{\rm MS}$ scheme fixes
$Z_2$ and $Z_m$ in order to subtract the contributions
$\sim \frac{1}{\epsilon}-\gamma+\ln 4\pi$ in Eq.~(\ref{sigma}).
The renormalized propagators $S^R$ in the on-shell (o.s.) and 
$\overline{\rm MS}$ schemes are then expressed in terms of pole and
$\overline{\rm MS}$ masses as follows:
\begin{equation}\label{sr}
S^R_{\rm o.s.}(p)\simeq \frac{i}{\cancel p-m_{\rm pole}}\ \ ,\ \  
S^R_{\overline{\rm MS}}\simeq \frac{i}{\cancel p-m_{\overline{\rm MS}}-(A-B)
m_{\overline{\rm MS}}}.\end{equation}
In Eq.~(\ref{sr}) $m_{\rm pole}$ is the pole
of the propagator after renormalization, which is in agreement with
the intuitive notion of the mass of a particle, whereas $m_{\overline{\rm MS}}$
may be quite far from the pole.
The relation between pole and $\overline{\rm MS}$ masses was calculated
up to four loops \cite{mspole} and reads, for top quarks:
\begin{equation}\label{polems}
m_{t,\rm pole}=\bar m_t(\bar m_t)
\left[1+0.42~\alpha_S+0.83~\alpha_S^2+2.37~\alpha_S^3+
(8.49\pm 0.25)~\alpha_S^4+{\cal O}(\alpha_S^5)\right].
\end{equation}
The last term in (\ref{polems}) yields
an uncertainty of about 200~MeV on the pole-$\overline{\rm MS}$
mass relation \cite{mspole}.

The self energy, when expressed in terms 
of the pole mass, is affected by infrared renormalons \cite{beneke}, i.e.
the coefficients of $\alpha_S^n$ grow factorially:
\begin{equation}\label{fact}
\Sigma^R(m_{\rm pole},m_{\rm pole})\approx m_{\rm pole}\ \sum_n\ \alpha_S^{n+1}\  
(2b_0)^n\ n!.
\end{equation}
Due to Eq.~(\ref{fact}), the pole mass gets corrections
$\Delta m_{\rm pole}\simeq {\cal O}(\Lambda_{\rm QCD})$, 
the so-called renormalon ambiguity.
The $\overline{\rm MS}$ mass is renormalon-free, 
but it is not a threshold
mass, as it exhibits corrections $(\alpha_S/v)^k$, 
$v$ being the top velocity, that are quite large for small $v$.

By using the recent computation \cite{mspole} to fit 
the ${\cal O}(\alpha_S^4)$ coefficient of the renormalon calculation in
\cite{beneke} and extrapolating the result to predict also the
higher-order terms, one can find that the renormalon ambiguity on the
pole mass is even below 100 MeV \cite{bns}.
This result, along with the good convergence of the expansion
(\ref{polems}), makes the top-quark pole mass a reliable quantity.

The MSR mass \cite{hoang} 
depends on a parameter $R$,
which could be, e.g., a factorization scale,  and
tries to interpolate between pole and $\overline{\rm MS}$ 
masses, i.e. $m^{\rm MSR}(R)\to m_{\rm pole}$ for
$R\to 0$ and $m^{\rm MSR}(R)\to \bar m(\bar m)$ for 
$R\to \bar m(\bar m)$.
Pole and MSR masses differ by a counterterm, i.e.
$m_{\rm pole}=m^{\rm MSR}(R,\mu)+\delta m(R,\mu)$,
where the $\mu$-dependence of $m^{\rm MSR}(R,\mu)$ follows
renormalization group equations.
The MSR mass is
typically employed in the context of Soft Collinear Effective
Theories (SCET). 
\section{Interpretation of the experimental results}
Standard experimental measurements
are carried out by using Monte Carlo simulations:
since parton showers are not exact QCD calculations, 
the interpretation
of the measured mass in terms of any field-theory mass definition
is not straightforward and one often calls it `Monte Carlo
mass'.
However, since these
measurements are done by reconstructing top-decay  ($t\to bW$)
observables, with
on-shell top quarks, the extracted
mass must mimic the pole mass, which is, by definition,
the pole of the renormalized propagator.
Such a simple picture is spoiled by the lack of
higher-order corrections, as
standard parton showers
are matched to the tree-level matrix element
\cite{corsey} and do not fully contain one-loop and
width ($\Gamma_t$) effects, or by 
colour-reconnection effects. In fact, in the 
Monte Carlo hadronization models, it may happen that, for
few events, 
the $b$ quark in top decay forms a $B$ meson with an antiquark from
the initial state.
Much work has therefore been undertaken to
estimate the uncertainty on the identification of
the measured mass with the pole mass.
 
As for NLO corrections,
in the aMC@NLO code NLO top decays are implemented for
single-top events \cite{rikk}; in $t\bar t$ production, 
the decays are on shell, but 
spin correlations and part of the off-shell contributions are
included via MadSpin \cite{madspin}.
In POWHEG, NLO top decays have been lately implemented 
\cite{pow}, accounting for $\Gamma_t$ effects 
in different approximations.

\begin{figure}[t!]
\centerline{\resizebox{0.472\textwidth}{!}
{\includegraphics{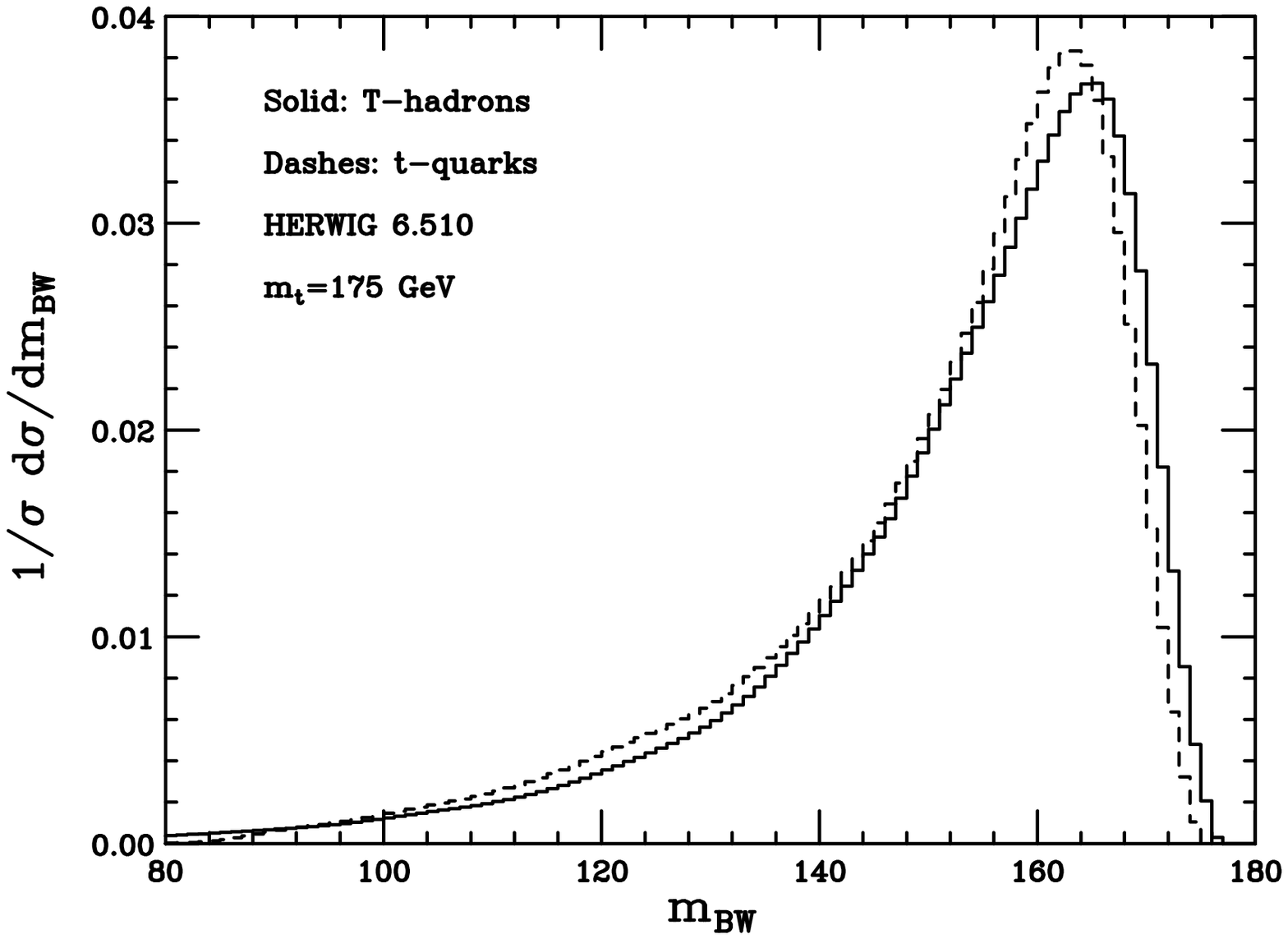}}\hspace{0.1cm}
\resizebox{0.472\textwidth}{!}{\includegraphics{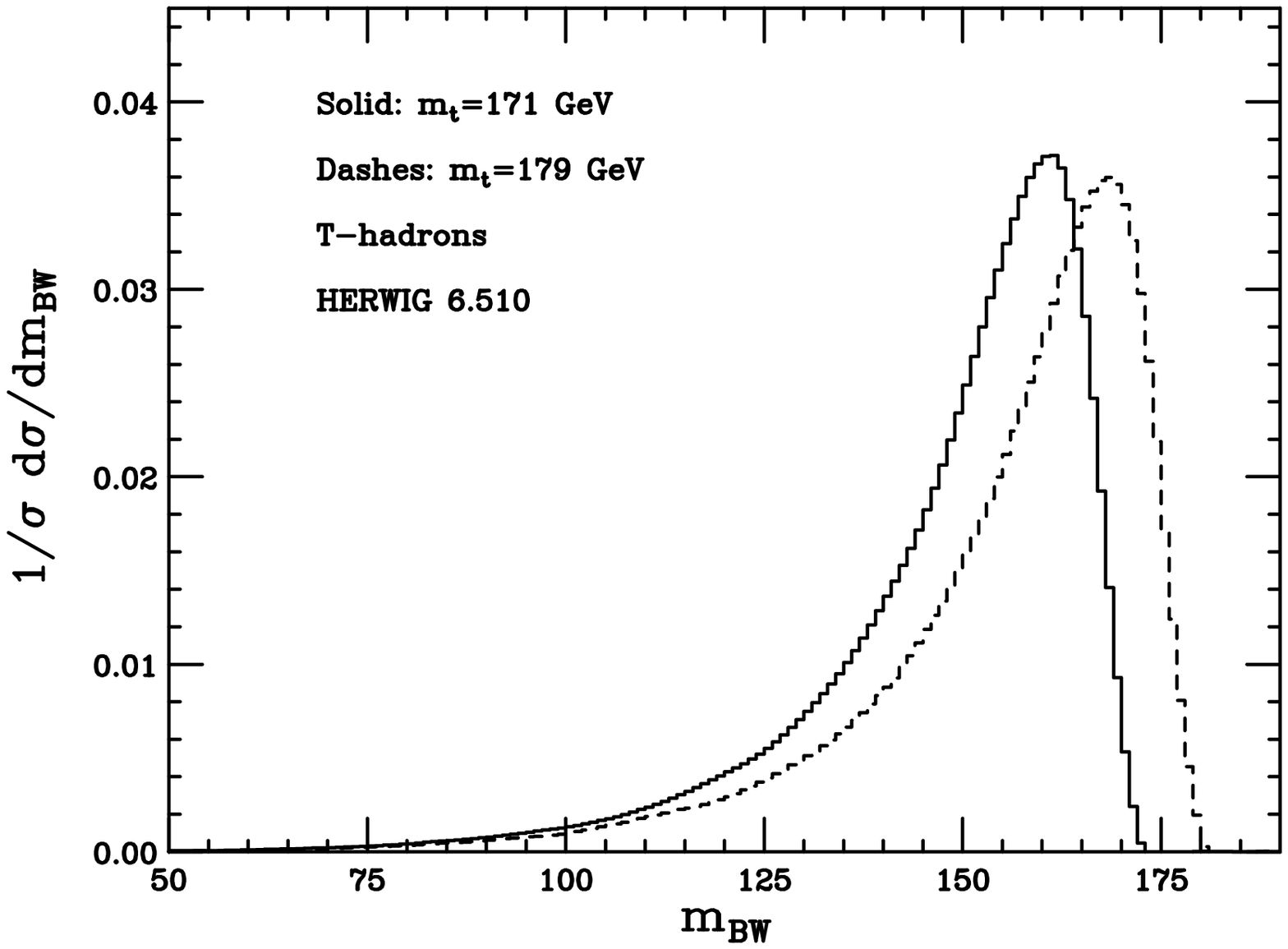}}}
\caption{$BW$ invariant mass in $e^+e^-\to t\bar t$
events at 1 TeV. Left: comparison for $T$-hadron
(solid) and standard top events (dashes) with $m_t=175$~GeV.
Right: $m_{BW}$ for $T$ hadrons and different values of $m_t$.}
\label{fig1}
\end{figure}

As far  as colour reconnection is concerned, Ref.~\cite{spyros} investigates
it in the framework of the Lund string model, tuned
to charged-particle multiplicity or
transverse momentum data. It is found
that the treatment of colour reconnection
can lead to an uncertainty on the top mass
within 200 and 500 MeV, according to the model which
is chosen. 
Colour connection in the HERWIG cluster model
was instead tackled in \cite{corc}, where top quarks
were forced to hadronize in top mesons and decay according
to the spectator model. In this way, top quarks must
form colour singlets with light quarks; also, by
using lattice-based methods, Heavy Quark Effective Theory
or Non Relativistic QCD, one can precisely
connect a meson mass to a well-defined
quark-mass definition.
Such a study does not aim at detecting $T$-hadrons, but rather,
by comparing observables in standard
$t\bar t$ samples and in $T$-hadron events, it
may shed light
on the non-perturbative uncertainty on $m_t$.
In Fig.~\ref{fig1} I present the $BW$ invariant mass
distribution, $B$ being a $b$-flavoured hadron in top decay,
for $e^+e^-\to t\bar t$ collisions
at 1 TeV, in the dilepton channel, by using the HERWIG 6 event generator.
If top quarks hadronize before decaying,
$m_{BW}$ is shifted towards higher values (Fig.~\ref{fig1}, left), 
with respect to standard top decays;
in fact, in $T$ decays, the $b$ quark  likely
forms with the spectator quark a cluster of small invariant mass, decaying
into a $B$ meson, plus soft hadrons, e.g. pions. Therefore, 
$m_{BW}$ tends to be closer to the kinematic limit, given by the mass of
the $T$-hadron. Figure~\ref{fig1} (right) presents the
$m_{BW}$ just for $T$ hadrons and 
$m_t=171$ and 179 GeV.
Work is in progress to quantify the discrepancies
in Fig.~\ref{fig1} in terms of an uncertainty on $m_t$
and its interpretation as a pole mass. 
\par
In the SCET framework \cite{hoang} the
jet mass plays the role of a MSR mass, with $R=\Gamma_t$ 
in the case of $e^+e^-\to t\bar t$ events, and 
can be related at NLO to the pole mass as follows:
\begin{equation}
m_J(\mu)=m_{\rm pole}-\frac{\alpha_S(\mu)C_F\Gamma_t}{\pi}
\left(\ln\frac{\mu}{\Gamma_t}+\frac{3}{2}\right)+{\cal O}(\alpha_S^2).
\end{equation}
One then assumes \cite{hoang1} that the measured $m_t$
is a jet mass at the scale of shower
cutoff $Q_0\simeq 1$~GeV, yielding an
uncertainty $m_{\rm pole}-m_J(Q_0)\simeq$~200 MeV.
Within SCET, one may compare resummed distributions,
such as the thrust in $e^+e^-\to t\bar t$, using 
the MSR mass for $R\sim {\cal O}(1~{\rm GeV})$, 
with Monte Carlo spectra and calibrate the Monte Carlo mass to reproduce the
SCET prediction \cite{hoang2}.

Another approach was suggested 
in \cite{moch1}: 
one first performs a simultaneous fit of the Monte Carlo mass and of
a given observable, such as total or differential cross sections,
and then compares with a (N)NLO calculation for the
same quantity, using, e.g., the pole mass.
The conclusion of Ref.~\cite{moch1} is that
the uncertainty on this calibration is roughly 2 GeV.

As for the alternative measurements,
a typical example is the total $t\bar t$
cross section, which was calculated to NNLO+NNLL accuracy in
\cite{alex}, and allows a
direct extraction of the pole mass \cite{sigmacms,sigmaatl}:
$m_t=\left(172.9^{+2.5}_{-2.6}\right)$~GeV (ATLAS) and
$m_t=\left(173.6^{+1.7}_{-1.8}\right)$~GeV (CMS), combining 7 and 8 TeV data.
In principle, even this extraction depends on the
use of event generators to evaluate the acceptance, but nevertheless
both ATLAS and CMS found a mild dependence on the mass implemented in
the Monte Carlo code.
Reference~\cite{dowling} investigated NNLO total and NLO differential
cross sections in terms of the $\overline{\rm MS}$ mass
and obtained an overall milder scale dependence; the recent
calculation of NNLO differential spectra \cite{mitov}
should shade more light on this finding.

The top pole mass was also determined from the $t\bar t+1$~jet cross
section, following \cite{ttj}, which has a stronger
dependence on $m_t$ with respect to the inclusive cross section.
ATLAS performed this measurement by using
POWHEG, matched to PYTHIA, taking care
of unfolding detector, hadronization and shower effects,
in such a way to recover the partonic $t\bar tj$ result.
The result is
$m_{t,{\rm pole}} =\left[173.7 \pm 1.5~{\rm (stat.)} \pm 
1.4~{\rm (syst.)}^{+1.0}_{-0.5}~({\rm th.})\right]$~GeV;
the impact of the Monte Carlo input mass in the evaluation of the 
acceptance is negligible.

Other observables, which have been lately proposed,
are the peak of the $b$-jet energy spectrum, 
the $b$-jet+lepton invariant mass $m_{b\ell}$ and a few
distribution endpoints. 
The general feature of these measurements is that 
they rely on the kinematic reconstruction of
top-decay final states and hence, once again, the extracted mass must
be close to the pole mass.
In detail, the $b$-jet energy measurement \cite{bj}
exhibits the property \cite{roberto}
that the position of the peak is independent of the boost 
from the top to the laboratory frame, as well as of the
production mechanism. The experimental measurement yields
$m_t=\left[172.29\pm~1.17~({\rm stat.})\pm~2.66~({\rm syst.})\right]$~GeV
at 8 TeV: however,
the invariance of the peak position is only valid at LO
and for inclusive spectra, and therefore
it will be interesting updating the analysis \cite{bj} by using
NLO top decays.

The $m_{b\ell}$ spectrum was used by CMS to reconstruct 
$m_t$ in the dilepton channel: by comparing with
the MadGraph+PYTHIA (LO) simulation, $m_t=(172.3\pm 1.3)$~GeV
was found \cite{mbl}.
Nevertheless, the NLO calculation of $m_{bl}$ \cite{melnikov},
employing the pole mass, is available and 
exhibits some discrepancies with respect to LO parton showers
\cite{corc,mescia}:  
an extension to NLO is thus mandatory.

Reference~\cite{end} measures $m_t$ from the endpoints of the
$m_{b\ell}$, $\mu_{bb}$ and $\mu_{\ell\ell}$ spectra, 
where $\mu_{bb}$ and $\mu_{\ell\ell}$
are related to the $bb$ and $\ell\ell$ invariant masses.
This analysis minimizes the Monte Carlo systematics,
since the $b$-jet can be calibrated 
directly from data; the leftover Monte Carlo uncertainties 
are mostly due to the assumption
that the $t$ and $\bar t$ decay chains are independent of 
colour reconnection.
The result, based on LO kinematic relations, is
$m_t=\left[173.9\pm 0.9 ({\rm stat.})^{+1.7}_{-2.1}({\rm syst.})\right]$~GeV;
comparing the data with the recent implementation of POWHEG \cite{pow},
accounting for some interference effects,
can therefore give some insight on the uncertainty due to higher-order
corrections.

\section{Conclusions}
I discussed the interpretation of the top mass results
at the LHC:
measurements relying on the reconstruction of top-decay products
yield results close to the top-quark pole mass, 
with an uncertainty due to the approximations in the computations
used for comparison, namely missing loop and width corrections
and colour-reconnection effects.
Using the recent calculation of the relation between pole and 
$\overline{\rm MS}$ masses, the renormalon ambiguity 
can be proved to be smaller than 100 MeV, thus making the pole mass
a suitable quantity.
Work has been done to quantify the
uncertainty on the interpretation of the measured mass as a pole mass,
by using Soft Collinear Effective Theories or simulating
fictitious top-hadron states.
Alternative measurements, based on the extraction
from the cross sections of $t\bar t$ and $t\bar t j$ events, yield 
the pole mass, up to small acceptance and hadronization 
corrections. Other strategies, using energy peaks, endpoints and
$m_{b\ell}$ look promising and worthwhile
to be pursued at 13 TeV, thanks to the higher statistics and the
late implementation of NLO top decays in shower generators.


\begin{thebibliography}{99}
\bibitem{degrassi}
G. Degrassi, S. Di Vita, J. Elias-Miro, J. R. Espinosa, G. F. Giudice, 
G. Isidori and A. Strumia, JHEP 1208 (2012) 098.
\bibitem{wave}
ATLAS and CDF and CMS and D0 Collaborations,  arXiv:1403.4427 [hep-ex].
\bibitem{atlas1}
ATLAS Collaboration, Eur. Phys. J. C75 (2015) 158.
\bibitem{cms1}
CMS Collaboration, arXiv:1509.04044 [hep-ex].
\bibitem{herwig6}
G. Corcella et al., JHEP 0101 (2001) 010.
\bibitem{pythia6}
T. Sj\"ostrand, S. Mrenna and P.Z. Skands,
JHEP 0605 (2006) 026.
\bibitem{mcnlo}
J. Alwall et al., JHEP 1407 (2014) 079.
\bibitem{powheg}
S. Alioli, P. Nason, C. Oleari and E. Re, JHEP 1006 (2010).
\bibitem{mspole}
P. Marquard, A.V. Smirnov, V.A. Smirnov and M. Steinhauser, 
Phys. Rev. Lett. 114 (2015) 142002.
\bibitem{beneke}
M. Beneke and V.N. Braun, Nucl. Phys. B426 (1994) 301.
\bibitem{bns}
P. Nason,  arXiv:1602.00443 [hep-ph].
\bibitem{hoang}
S. Fleming, A.H. Hoang, S. Mantry, I.W. Stewart,
Phys. Rev. D77 (2008) 114003.
\bibitem{spyros}	
S. Argyropoulos and T. Sj\"ostrand, JHEP 1411 (2014) 043.
\bibitem{corsey}
G. Corcella and M.H. Seymour, Phys. Lett. B442 (1998) 417.
\bibitem{rikk}
A.S. Papanastasiou, R. Frederix, S. Frixione, V. Hirschi and F. Maltoni,
Phys. Lett. B726 (2013) 223.
\bibitem{madspin}
P. Artoisenet, R. Frederix, O. Mattelaer and R. Rietkerk,
JHEP 1303 (2013) 015.
\bibitem{pow}
J.M. Campbell, R.K. Ellis, P. Nason and E. Re, JHEP 1504 (2015) 114.
\bibitem{corc}
G. Corcella, EPJ Web Conf. 80 (2014) 00019.
\bibitem{hoang1}
A.H. Hoang and I.W. Stewart,
Nucl. Phys. Proc. Suppl. 185 (2008) 220.
\bibitem{hoang2}
A.H. Hoang,  arXiv:1412.3649 [hep-ph]. 
\bibitem{moch1}
J. Kieseler, K. Lipka and S.-O. Moch, arXiv:1511.00841 [hep-ph].
\bibitem{alex}
M. Czakon, P. Fiedler and A.D. Mitov, Phys. Rev. Lett. 110 (2013) 252004.
\bibitem{sigmacms}
CMS Collaboration, CMS-PAS-TOP-13-004.
\bibitem{sigmaatl}
ATLAS Collaboration, Eur. Phys. J. C74 (2015) 3109.
\bibitem{dowling}
M. Dowling and S.-O. Moch, Eur. Phys. J. C74 (2014) 3167.
\bibitem{mitov}
M. Czakon, D. Heymes and A.D. Mitov, Phys. Rev. Lett. 116 (2016) 082003.
\bibitem{ttj}
S. Alioli, P. Fernandez, J. Fuster, A. Irles 
S.-O. Moch, P. Uwer and M. Vos, Eur. Phys. J. C73 (2013) 2438.
\bibitem{atlttj}
ATLAS Collaboration, JHEP 1510 (2015) 121.
\bibitem{bj}
CMS Collaboration, CMS-PAS-TOP-15-002.
\bibitem{roberto}
K. Agashe, R. Franceschini and D. Kim, Phys. Rev. D88 (2013) 057701.
\bibitem{mbl}
CMS Collaboration, CMS-PAS-TOP-14-014.
\bibitem{melnikov}
S. Biswas, K. Melnikov and M. Schulze, JHEP 1008
(2010) 048.  
\bibitem{mescia}
G. Corcella and F. Mescia, Eur. Phys. J. C65 (2010) 171;
Erratum-ibid. C68 (2010) 687.
\bibitem{end}
CMS Collaboration, Eur. Phys. J. C73 (2013) 2494.
\end{thebibliography}
\end{document}